Chapter 8

# Enhanced Rank-Based Correlation Estimation Using Smoothed Wilcoxon Rank Scores 

Feridun Taşdan[1]

Rukiye Dağalp[2]

*Abstract*

*This article proposes an improved version of the Spearman rank correlation based on using Wilcoxon rank score function. A smoothed empirical cumulative distribution function (ecdf) computes the smoothed ranks and replaces the regular ranks in the Wilcoxon rank score function. The smoothed Wilcoxon rank scores are then used for estimation of the Spearman's correlation. The proposed approach is similar to the Spearman's rho ($\rho$) estimator which uses ranks of the random samples of $X$ and $Y$ but the proposed method improves Spearman's approach such as handling ties and gaining higher efficiency under monotone associations. A Wald type hypothesis test has been proposed for the new estimator $r_{sa}$ and the asymptotic properties are shown.*

## 1 INTRODUCTION

The rank based estimation and testing procedures for the correlation coefficient $\rho$ are commonly used in many data analysis problems. Especially Spearman's correlation (Spearman, 1904) and Pearson's correlation, (Pearson, 1920) coefficients are the leading methods of estimation the correlations between two dependent random variables. Pearson's traditional approach works well if the relationship is linear and the underlying joint distribution is normal. On the other hand, Spearman's approach is based on the ranks of the random samples and does not depend on the normality assumption or any other underlying distribution. Spearman's approach works well if the relationship is linear or monotonic as discussed and shown by (Santos et al.,2011), (Fujita et al., 2009), (Szekely et al.,2007) and (Reshef et al., 2011). Based on these studies, it has been shown that Spearman has some superior properties over the Pearson's correlation if the underlying conditions are not ideal such as having a monotonic association instead of a linear. Another traditional estimator Kendall's $\tau$ (Kendall,1938) has been included in the study as it is also a nonparametric approach. Hoeffding's D (Hoeffding, 1948) and a few more other are not included in this study in order to focus on smoothed ranks with respected to its traditional competitors under linear and monotonic associations. In the future, an extensive comparative study of correlation methods under non-monotonic associations would be

1   Department of Mathematics Western Illinous University, Macomb, IL 61455
2   Department of Statistics Ankara University, Ankara, Turkey.

considered, and there could be an improved version of the smoothed ranks could be introduced in the future for non-monotonic cases. Moreover, smoothing ranks and using them in parameter estimations or in regression methods have been proposed by several authors; (Heller, 2007), (Heller at al., 2012), (Lin and Peng, 2013), (Serfling, 1984), (Tasdan and Yeniay, 2014), and (Tasdan, 2018).

The figures below depict the cases of linear, nonlinear monotonic and non-monotonic nonlinear relationships.

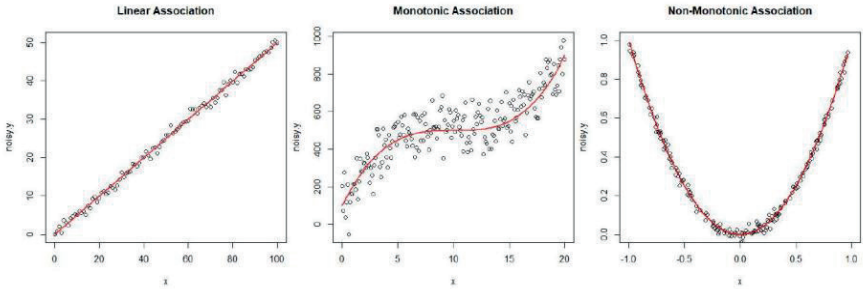

**Figure 1:** Examples of linear, monotone and non-monotone relationships
In Section 2 traditional correlation estimations are reviewed. In Section 3 score-based estimation of the correlation coefficient $\rho$ is discussed. Moreover, the relationship between the Spearman's rank-based correlation estimation and the Wilcoxon score-based correlation estimations are shown. In Section 4 it is shown that a non-traditional approach of smoothed ranks is used in the Wilcoxon's score function to estimate the correlation coefficient. In addition, Wald type hypothesis testing option has been proposed. In Section 5 Monte Carlo simulation study has been performed to show small and large sample properties under several bivariate distributions. Final conclusion about the proposed estimator and the results of the Monte Carlo simulations that study is presented in Section 6.

## 2 PEARSON AND SPEARMAN'S CORRELATION ESTIMATIONS

The measure of association between random variables $X$ and $Y$ is defined as $\rho$ which is called the population correlation coefficient. Pearson's $\rho$ definition which is given by (Pearson, 1920) can be written as

$$\rho = \frac{E[(X - \mu_x)(Y - \mu_y)]}{\sigma_x \sigma_y} = \frac{\text{Cov}(X, Y)}{\sigma_x \sigma_y} \#(1)$$

where $-1 \leq \rho \leq +1$ and $\text{Cov}(X, Y) = E[XY] - E[X]E[Y]$, which is defined as covariance between $X$ and $Y$. It should be noted that if $\rho = 0$ (independence of X and Y ) implies $Cov(X,Y) = 0$ but the opposite is not true.

First we find the method of moment estimator for Pearson's $\rho$, let $(X_1, Y_1), (X_2, Y_2), .., (X_n, Y_n)$ be a random sample from a bivariate continuous cdf of $F(X,Y)$ and let r be the estimated sample correlation coefficient. The method of moment estimator of

$$Cov(X,Y) = E[(X - \mu_x)(Y - \mu_y)] \#(2)$$

can be defined as

$$\widehat{Cov(X,Y)} = \frac{1}{n} \sum_{i=1}^{n} (X_i - \bar{X})(Y_i - \bar{Y}) \#(3)$$

The $\sigma_x$ can be estimated with

$$\widehat{\sigma_x} = \sqrt{\frac{1}{n} \sum_{i=1}^{n} (X_i - \bar{X})^2} \#(4)$$

and $\sigma_y$ can be estimated with

$$\widehat{\sigma_y} = \sqrt{\frac{1}{n} \sum_{i=1}^{n} (Y_i - \bar{Y})^2} \#(5)$$

Thus, the Pearson's method of moment correlation coefficient estimate can be written as

$$r_p = \frac{\sum_{i=1}^{n} (X_i - \bar{X})(Y_i - \bar{Y})}{\sqrt{\sum_{i=1}^{n} (X_i - \bar{X})^2} \sqrt{\sum_{i=1}^{n} (Y_i - \bar{Y})^2}} \#(6)$$

The result above can be further simplified to the version below,

$$r_p = \frac{\sum_{i=1}^{n} X_i Y_i - \frac{1}{n} (\sum_{i=1}^{n} X_i)(\sum_{i=1}^{n} Y_i)}{\sqrt{\left[\sum_{i=1}^{n} X_i^2 - \frac{1}{n}\left(\sum_{i=1}^{n} X_i\right)^2\right]\left[\sum_{i=1}^{n} X_i^2 - \frac{1}{n}\left(\sum_{i=1}^{n} X_i\right)^2\right]}} \#(7)$$

A nonparametric (rank based) alternative of Pearson's correlation coefficient is proposed by Spearman (Spearman,1904). Spearman's $\rho$ estimator uses the ranks of random samples of $X$ and $Y$ instead of the original observations. Let $R(X_i)$ be the rank of $ith$ observation $X_i$ for $i = 1, \ldots, n$. Similarly, $Y_i$ is replaced by $R(Y_i)$. By substituting the ranks $R(X_i)$ and $R(Y_i)$ in the Pearson's correlation coefficient formula in Eq(6), we obtain the Spearman's correlation coefficient estimate below

$$r_s = \frac{\sum_{i=1}^{n} \left(R(X_i) - \overline{R(X_i)}\right)\left(R(Y_i) - \overline{R(Y_i)}\right)}{\sqrt{\sum_{i=1}^{n} \left(R(X_i) - \overline{R(X_i)}\right)^2 \sum_{i=1}^{n} \left(R(Y_i) - \overline{R(Y_i)}\right)^2}} \#(8)$$

We should note that the ranks $R(X_i)$ (also $R(Y_i)$ ) are uniformly distributed for the integers on $i = 1, \ldots, n$. Also, sum of the ranks can be written as

$$\sum_{i=1}^{n} R(X_i) = \sum_{i=1}^{n} R(Y_i) = n(n+1)/2 \#(9)$$

for $i = 1, \ldots, n$. Thus, $\overline{R(X_i)} = \overline{R(Y_i)} = \frac{n+1}{2}$. Moreover, it can be shown that

$$E[R(X_i)] = (n+1)/2 \text{ and } nV[R(X_i)] = \sum_{i=1}^{n} \left(R(X_i) - \overline{R(X_i)}\right)^2 = \frac{n(n^2-1)}{12} \#(10)$$

So, we can use these results to obtain much simpler version of $r_s$ as described below;

$$r_s = \frac{\sum_{i=1}^{n} \left(R(X_i) - \frac{n+1}{2}\right)\left(R(Y_i) - \frac{n+1}{2}\right)}{\frac{n(n^2-1)}{12}} \#(11)$$

Yet, there is another version of the Spearman's correlation coefficient has been developed in the literature or used in teachings of the correlation coefficients. So, let $D_i = R(X_i) - R(Y_i)$ for $i = 1, \ldots, n$. It is also true that

$$D_i = \left[R(X_i) - \overline{R(X_i)}\right] - \left[R(Y_i) - \overline{R(Y_i)}\right] \#(12)$$

since $\overline{R(X_i)} = \overline{R(Y_i)} = \frac{n+1}{2}$. Thus, it can be written that

$$\sum_{i=1}^{n} D_i^2 = \sum_{i=1}^{n} \left[ R(X_i) - \overline{R(X_i)} \right] - \left[ R(Y_i) - \overline{R(Y_i)} \right]^2 \#(13)$$

$$= \sum_{i=1}^{n} \left( R(X_i) - \overline{R(X_i)} \right)^2 + \sum_{i=1}^{n} \left( R(Y_i) - \overline{R(Y_i)} \right)^2 \#(13)$$

$$- 2 \sum_{i=1}^{n} \left( R(X_i) - \overline{R(X_i)} \right) \left( R(Y_i) - \overline{R(Y_i)} \right) \#(13)$$

$$\sum_{i=1}^{n} D_i^2 = \frac{n(n^2 - 1)}{12} + \frac{n(n^2 - 1)}{12} - 2 \sum_{i=1}^{n} \left( R(X_i) - \overline{R(X_i)} \right) \left( R(Y_i) - \overline{R(Y_i)} \right) \#(13)$$

If the last expression on the right of the equation is isolated, we can obtain

$$\sum_{i=1}^{n} \left( R(X_i) - \overline{R(X_i)} \right) \left( R(Y_i) - \overline{R(Y_i)} \right) = \frac{n(n^2 - 1)}{12} - \frac{\sum_{i=1}^{n} D_i^2}{2} \#(14)$$

So, the numerator in Eq (8) is replaced with the above result, and the denominator terms (square root of variances) are replaced with $\sum_{i=1}^{n} \left( R(X_i) - \overline{R(X_i)} \right)^2 = \frac{n(n^2-1)}{12}$, then a further simplification gives us yet another definition of $r_s$ which can algebraically be reduced to

$$r_s = 1 - \frac{6 \sum_{i=1}^{n} D_i^2}{n(n^2 - 1)} \#(15)$$

where $D_i = R(X_i) - R(Y_i)$. If some ties exist in the samples, a small estimation error might occur in the last formula but a few ties can be tolerated without a significant difference. The average method can be used to break the ties if necessary. All three versions of $r_s$ produce the same result but each has somewhat different computational difficulties.

## 3 SPEARMAN'S RANK CORRELATION BASED ON GENERAL SCORE FUNCTIONS

Spearman's $r_s$ estimator of the correlation parameter $\rho$ can be computed with using a general score function $\varphi(u)$ where the score function must be nondecreasing function defined on the interval (0,1) such that $\int_0^1 \varphi^2(u) du < \infty$. (Hettmansperger, 1984) and (Hettmansperger and McKean, 2011) defines the correlation coefficient estimator based on a general score function $\varphi(u)$ as

$$r_a = \frac{1}{s_a^2} \sum_{i=1}^{n} a(R(X_i))a(R(Y_i)) \#(16)$$

where $a(i) = \varphi(i/n + 1)$ and $s_a^2 = \sum_{i=1}^{n} a^2(i)$ for $i = 1, \ldots, n$. The score functions $\varphi(u)$ can be standardized so that it satisfies $\int_0^1 \varphi(u)du = 0$ and $\int_0^1 \varphi^2(u)du = 1$. Moreover, $E(r_a)$ and $Var(r_a)$ can be derived with the following theorem.

**Theorem 1.** Let $(X_1, Y_1), \ldots, (X_n, Y_n)$ be a random sample from a bivariate continuous distribution function $F(X,Y)$. Under the null hypothesis of $H_0: \rho = 0$ ($F(X,Y) = F_x(X)F_y(Y)$ or independence of X and Y), the test statistics $r_a$ satisfies the following properties:
a) $E[r_a] = 0$
b) $Var[r_a] = 1/(n-1)$

**Proof.** Since it is assumed that the scores are generated by $a(i) = \varphi(i/n + 1)$ for $i = 1, \ldots, n$ and $a(1) \leq a(2) \leq \cdots \leq a(n)$ based on the assumptions that the score function $\varphi(u)$ is nondecreasing. The score function is also centered so that $\int_0^1 \varphi(u)du = 0$ and $\int_0^1 \varphi^2(u)du = 1$. By using the Reimann-Sum approximation on the integral, it can be shown that $\int_0^1 \varphi(u)du \doteq \sum_{i=1}^{n} \varphi(i/(n+1)) \frac{1}{n} = \sum_{i=1}^{n} a(i) \frac{1}{n} \approx 0$. Also, $\int_0^1 \varphi^2(u)du \doteq \sum_{i=1}^{n} \varphi^2(i/(n+1)) \frac{1}{n} = \sum_{i=1}^{n} a^2(i) \frac{1}{n} = \frac{s_a^2}{n} \approx 1$ can be approximated in a similar fashion.

It is known that the ranks, $R(X_i)$ (also $R(Y_i)$ ), is uniformly distributed for the integers on $i = 1, \ldots, n$. Thus, $P[(R(X_i) = k] = 1/n$, where k is the rank of ith observation. So, each rank of $X_i$ is equally likely distributed with a probability of $1/n$. Therefore, from the expected value of $a(R(X_i))$ it can be found that $E[a(R(X_i))] = \sum_{i=1}^{n} a(i) \frac{1}{n} = 0$ since $\sum_{i=1}^{n} a(R(X_i)) = 0$. Similarly, $E[a(R(Y_i))] = \sum_{i=1}^{n} a(i) \frac{1}{n} = 0$. When above results are substituted into the expression of $r_a$ as defined in Eq 16, it shows $E[r_a] = 0$.

To show part b, the variance definition of $V[r_a] = E[(r_a)^2] - (E[r_a])^2$ can be used for the estimator $r_a$. In the first part of the proof (part-a), it has been shown that $E[r_a] = 0$. Thus, the variance expression can be simplified to $V[r_a] = E[(r_a)^2]$.

Assuming under $H_0: \rho = 0$,

$$E[(r_a)^2] = E\left\{\left[\frac{1}{s_a^2}\sum_{i=1}^{n} a(R(X_i))a(R(Y_i))\right]^2\right\}$$

$$= \frac{1}{s_a^4}\sum_{i=1}^{n} E[a(R(X_i))a(R(Y_i))]\sum_{j=1}^{n} E\left[a\left(R(X_j)\right)a\left(R(Y_j)\right)\right]$$

$$= \frac{1}{s_a^4}\sum_{i=1}^{n}\sum_{j=1}^{n} E[a(R(X_i))a(R(Y_i))]E\left[a\left(R(X_j)\right)a\left(R(Y_j)\right)\right]$$

$$= \frac{1}{s_a^4}\sum_{i=1}^{n}\sum_{j=1}^{n} E\left[a(R(X_i))a\left(R(X_j)\right)\right] E\left[a(R(Y_i))a\left(R(Y_j)\right)\right]$$

To find the result of the expectations in the last expression, there are two cases ($i = j$ and $i \neq j$) that must be considered separately. For $i = j$,

$$E\left[a(R(X_i))a\left(R(X_j)\right)\right] = E[a^2(X_i)] = \sum_{i=1}^{n} a^2(X_i)\frac{1}{n} = \frac{1}{n}s_a^2$$

Similarly, it can be found that

$$E\left[a(R(Y_i))a\left(R(Y_j)\right)\right] = \frac{1}{n}s_a^2$$

For $i \neq j$, it can be written

$$E\left[a(R(X_i))a\left(R(X_j)\right)\right] = \sum_{i=1}^{n}\sum_{j=1}^{n} a(R(X_i))a\left(R(X_j)\right)\frac{1}{n(n-1)}$$

$$= \frac{1}{n(n-1)}s_a^2$$

The last result can be applied to $E\left[a(R(Y_i))a\left(R(Y_j)\right)\right] = \frac{1}{n(n-1)}s_a^2$. So, by substituting both cases of $i = j$ and $i \neq j$,

$$E[(r_a)^2] = \frac{1}{s_a^4} \sum_{i=1}^{n} \sum_{j=1}^{n} E\left[a(R(X_i))a\left(R(X_j)\right)\right] E\left[a(R(Y_i))a\left(R(Y_j)\right)\right]$$

$$= \frac{1}{s_a^4} \left\{ \sum_{i=j=1}^{n} E[a^2(X_i)]E[a^2(Y_i)] + \sum_{i \neq j} \sum_{i \neq j} E\left[a(R(X_i))a\left(R(X_j)\right)\right] E\left[a(R(Y_i))a\left(R(Y_j)\right)\right] \right\}$$

$$= \frac{1}{s_a^4} \left\{ \sum_{i=j=1}^{n} \frac{1}{n} s_a^2 \frac{1}{n} s_a^2 + \sum_{i \neq j} \frac{1}{n(n-1)} s_a^2 \frac{1}{n(n-1)} s_a^2 \right\}$$

$$= \frac{1}{s_a^4} \left[ \frac{n s_a^4}{n^2} + n(n-1) \frac{s_a^4}{n^2(n-1)^2} \right]$$

$$= \frac{1}{s_a^4} \left[ \frac{s_a^4}{n} + \frac{s_a^4}{n(n-1)} \right]$$

$$= \frac{1}{s_a^4} \left[ \frac{(n-1)s_a^4}{n(n-1)} + \frac{s_a^4}{n(n-1)} \right]$$

$$= \frac{1}{s_a^4} \frac{s_a^4}{n-1}$$

$$E[(r_a)^2] = \frac{1}{n-1}$$

Thus, using the results of $E[r_a] = 0$ and $E[(r_a)^2] = \frac{1}{n-1}$ as shown above, we prove that $V[r_a] = \frac{1}{n-1}$. The finite sample distribution of $r_a$ is not easy to find but it can be claimed that (by the central limit theorem) $z_a = \frac{r_a}{\sqrt{1/(n-1)}} = r_a\sqrt{n-1} \to N(0,1)$ as $n$ goes to $\infty$. For an asymptotic (large sample) $\alpha$ level test, it can be defined that reject $H_0: \rho = 0$ if $|z_a| > z_{\alpha/2}$.

For example, $\varphi(u)$ can be taken as $\varphi(u) = \sqrt{12}(u - 1/2)$ which is called the Wilcoxon's linear score function. It can be shown (with the following theorem) that Spearmen's rank correlation ($r_s$) is equal to to the $r_a$ which is defined in Eq(16) if Wilcoxon's linear score function is used to estimate the correlation coefficient of $r_a$.

**Theorem 2.** Let $(X_1, Y_1), \ldots, (X_n, Y_n)$ be a random sample from a bivariate continuous distribution function $F(X, Y)$. Also let $\varphi(u) = \sqrt{12}(u - 1/2)$ which is called the Wilcoxon's linear score function. Under the null hypothesis of $H_0: \rho = 0 (F(X, Y) = F_x(X)F_y(Y)$ or independence of $X$ and $Y$), the test statistics based on Wilcoxon's rank score $r_a$ is equal to the Spearman's $r_s$ with the same variance and expected value as defined in Theorem 1.

**Proof.** First, recall that

$$a(i) = \varphi[(i/n + 1)] = \sqrt{12}(i/(n + 1) - 1/2) \#(17)$$

which is called Wilcoxon's score function. Also, according to (Hettmansperger, 1984), $s_a^2$ is defined as

$$s_a^2 = \sum_{i=1}^{n} a^2(i)$$
$$= \sum_{i=1}^{n} \{\sqrt{12}[i/(n+1) - 1/2]\}^2$$
$$= \frac{12}{(n+1)^2}\left[\sum_{i=1}^{n} i^2 - (n+1)\sum_{i=1}^{n} i + \frac{n(n+1)^2}{4}\right]$$
$$= \frac{n(n-1)}{n+1}$$

So, we also let

$$a(R(X_i)) = \sqrt{12}[R(X_i)/(n+1) - 1/2] \#(18)$$

and similarly,

$$a(R(Y_i)) = \sqrt{12}[R(Y_i)/(n+1) - 1/2] \#(19)$$

We plug in above results into the $r_a$ which is defined in Eq (16). Then, we can find

$$r_a = \frac{\sum_{i=1}^{n}[(R(X_i) - (n+1)/2)(R(Y_i) - (n+1)/2)]}{n(n^2 - 1)/12} \#(20)$$

which is also defined as $r_s$ in Eq (11). So, the result of the Theorem 2 proves that Spearman's $r_s$ can also be estimated using a score based methods such as Wilcoxon's score function.

Yet there is another nonparametric estimator available to estimate $\rho$, which is called Kendall's $\tau$ estimator (Kendall, 1938). It still uses the ranks of the observations but it counts the concordant and discordant pairs of the observations in the pairs of random samples. An estimator of Kendall's $\tau$ can be derived by

$$r = \frac{(n_c - n_d)}{(n(n-1))/2} \#(21)$$

where $n_c$ shows the number of concordant pair and $n_d$ shows the number of discordant pairs.

If we compare Pearson against Kendall's and Spearman's estimators, Pearson's correlation coefficient measures how well the linear relationship between two variables are but in reality, there could be other ways that two variables can be correlated such as monotonic relationship which does not show relationship with a straight line. Monotonic relationships between two variables can show exponential or logistics distributions such as while one variable increases, the other increases or decreases consistently. Spearman and Kendall's correlation coefficients are basically a non-parametric way to investigate the monotonic relationships between two random variables with a continuous bivariate distribution function of $F(X,Y)$.

The Spearman correlation test assumes only that your data are a random sample. Spearman's correlation is calculated on the ranks of the observations and will therefore work with any type of data that can be ranked, including ordinal, interval, or ratio data. Because it based on ranks, it is less sensitive to outliers than the Pearson correlation test, and it is sometimes used to evaluate a correlation when outliers are present.

## 4 PROPOSED SMOOTHED RANKS BASED CORRELATION ESTIMATOR

First, recall that the general score function is defined as

$$a(i) = \varphi[(i/n+1)] = \sqrt{12}(i/(n+1) - 1/2)$$

which is based on Wilcoxon's linear score function. Then using regular ranks, we can write the score function as

$$a(R(X_i)) = \sqrt{12}[R(X_i)/(n+1) - 1/2] \#(22)$$

and similarly,

$$a(R(Y_i)) = \sqrt{12}[R(Y_i)/(n+1) - 1/2] \#(23)$$

As it was shown by Theorem 1 and Theorem 2, $r_a$ is equal to the Spearman's $r_s$ as defined in Eq(15). The proposed method replaces $R(X_i)$ and $R(Y_i)$ with a smoothed ranks and call them $\widehat{R(X_i)}$ and $\widehat{R(Y_j)}$, respectively. The proposed approach and the method described below has been introduced by (Tasdan, 2018).

In order to generate a rank set of a random sample of $X_1, X_2, .., X_n$, the empirical cdf of the random sample is defined as

$$F_n(x) = \frac{\#\{X_j \leq x\}}{n} = \frac{1}{n} \sum_{j=1}^{n} \mathbf{I}(X_j \leq x) \#(24)$$

where **I** is called the indicator function. This is a stepwise discrete function at each $X_i$. Then, the ranks of a random sample can be expressed as

$$R(X_i) = nF_n(X_i) = \sum_{j=1}^{n} \mathbf{I}(X_j \leq X_i) \#(25)$$

where $R(X_i)$ is the rank of the $i$ th observation. In a similar fashion,

$$R(Y_i) = nF_n(Y_i) = \sum_{j=1}^{n} \mathbf{I}(Y_j \leq Y_i) \#(26)$$

We now consider a smooth approximation to the indicator function by replacing I with a continuous distribution function $H(x/h)$, where the scale parameter $h$ is called the bandwidth or smoothing parameter in kernel smoothing applications and $h$ approaches zero as the sample size increases. The distribution function $H(x)$ is chosen to be a continuous, nondecreasing, and also bounded function that generates smoothed ranks as described below. The resulting smoothed empirical function can be written as

$$F_s(t) = \frac{1}{n} \sum_{j=1}^{n} H\left(\frac{t - X_j}{h}\right) \#(27)$$

By using this result, we can now define the smoothed ranks

$$\widehat{R(X_i)} = nF_s(X_i) = \sum_{j=1}^{n} H\left(\frac{X_i - X_j}{h}\right) \#(28)$$

where the notation $\widehat{R(X_i)}$ indicates the smoothed rank of the observation $X_i$. It is important to mention that the purpose of smoothing is not about estimating a density function but to use $H(x)$ function as a smooth approximation of the indicator function $I$. Therefore, the smoothed ranks can be generated from the $F_s(t)$ function. Moreover, it should be noted that when $x_i > X_j$, $H\left(\frac{x_i - X_j}{h}\right) \to 1$ when $h$ approaches zero as $n$ gets large.

Similarly, when $x_i < X_j$, $H\left(\frac{x_i - X_j}{h}\right) \to 0$ when $h$ approaches zero as $n$ gets large.

Also, it should be noted that

$$H_n(X_{(i)}) = \frac{i-1}{n} + \frac{X_{(i)} - V_i}{n(V_{i+1} - V_i)} \#(29)$$

where $X_{(i)}$ is the $i$ th order observation and $V_i$ is the midpoints between $X_i$ and $X_{i+1}$. Also, $V_1 = X_1 - (V_2 - X_1)$ and $V_{n+1} = X_n + (Z_n - V_n)$.

As a result of the smoothing and replacing the indicator function I with $H(x/h)$, we now obtain the smoothed ranks using smoothed distribution function of $F_s(X_i)$ and write them as

$$\widehat{R(X_i)} = nF_s(X_i) \#(30)$$

and

$$\widehat{R(Y_i)} = nF_s(Y_i) \#(31)$$

Then using these smoothed ranks, we rewrite the smoothed score function

$$a(\widehat{R(X_i)}) = \sqrt{12}[\widehat{R(X_i)}/(n+1) - 1/2] \#(32)$$

and similarly,

$$a(\widehat{R(Y_i)}) = \sqrt{12}[\widehat{R(Y_i)}/(n+1) - 1/2] \#(33)$$

We can plug in the new scores into $r_a$ as defined in Eq (16) to derive the new version of Spearman's correlation estimator based on the smoothed Wilcoxon score function,

$$r_{sa} = \frac{1}{s_a^2} \sum_{i=1}^{n} a(\widehat{R(X_i)}) a(\widehat{R(Y_i)}) \#(34)$$

where $s_a^2 = \sum_{i=1}^{n} a^2(i) = \frac{n(n-1)}{n+1}$ with using Wilcoxon's score function. A further simplification of the above results gives

$$r_{sa} = \frac{\sum_{i=1}^{n} \left[(\widehat{R(X_i)} - (n+1)/2)(\widehat{R(Y_i)} - (n+1)/2)\right]}{n(n^2 - 1)/12} \#(35)$$

Moreover, we can replace $R(X_i)/(n+1) = \frac{n}{n+1} F_n(X_i)$ by $H_n(X_{(i)})$ and $R(Y_i)/(n+1) = \frac{n}{n+1} F_n(Y_i)$ by $H_n(X_{(i)})$ and $H_n(Y_{(i)})$, respectively. Then,

$$a(\widehat{R(X_i)}) = \sqrt{12}[\widehat{H_n(X_i)} - 1/2] \#(36)$$

Similarly,

$$a(\widehat{R(Y_i)}) = \sqrt{12}[\widehat{H_n(Y_i)} - 1/2] \#(37)$$

Then we can plug in this result into $r_a$ and found the new version of the Spearman correlation coefficient estimator based on kernel function

$$r_{sa} = \frac{\sum_{i=1}^{n}\left[(H_n(X_{(i)}) - (n+1)/2)(H_n(Y_{(i)}) - (n+1)/2)\right]}{n(n^2-1)/12} \#(38)$$

## 4.1 BANDWIDTH SELECTION

It is well known that the selection of bandwidth $h$ is more important than the shape of the kernel function $H(x)$ as stated by (Sheather,2004) and (Silverman, 1986). As a result, an appropriate selection of a bandwidth $h$ must be considered in order to obtain a smoothed approximation of the indicator function. There are many bandwidth $h$ options available in the smoothing applications but the bandwidths suggested by (Silverman, 1986), (Sheather and Jones, 1991), and Bowman (1984) are options to be considered because of their common usage, high performance and also availability in R software. The bandwidth suggested by (Silverman, 1986) is also called the "rule of thumb" bandwidth approach in the literature. It has a numerical formula that equals to $h = 0.9\hat{\sigma}n^{-0.20}$, where $\hat{\sigma} = \min\{s, IQR/1.349\}$. (Bowman, 1984) suggests a least squares cross-validation approach and it is also called an unbiased cross-validation smoother in the literature. Finally, Sheather-Jones's plug-in bandwidth approach is also commonly used and considered to be a good performer as suggested by (Simonoff, 1996). The last two options have no closed form formulas. (Tasdan and Yeniay, 2014) discussed these three bandwidth options for the smoothed version of the KolmogorovSmirnov statistic, and has shown all three bandwidth options performed similarly in estimating the shift parameter for location problems. Moreover, we also considered the smoothing parameter $h = \hat{\sigma}n^{-0.26}$ as suggested by (Heller, 2007). It also satisfies $nh \to \infty$ and $nh^4 \to 0$ conditions in order to have an optimal rate of convergence. The $\hat{\sigma}$ is the estimated pooled standard deviation from the data. For robustness purposes, we consider using MAD (Median Absolute Deviation) to estimate $\sigma$ instead of a regular standard deviation approach. Comparative study of these four different bandwidth parameters will be discussed in detail later in Section 5.

## 5 MONTE CARLO SIMULATION STUDY

In this section, Monte Carlo Simulation has been performed and the Mean Square Error (MSE) of the $\rho$ parameter for each estimator $r$ is computed. The MSE is formulated as $MSE(r) = (r - \rho)^2 + V(r)$ which is found by average of bias $^2$ and adding the variance of correlation estimator(r). First, two random samples of size $n = 50$ are generated from a bivariate normal distribution with parameters $\mu_1 = 2, \mu_2 = 4, \sigma_1 = 1, \sigma_2 = 1$ and with the correlation parameter $\rho$ which is set to the sequence of $\rho = \{0, 0.02, 0.04, \ldots, 1\}$.

For each parameter value of the $\rho$, Person's $r_p$, Spearman's $r_s$, Kendall's $r_k$ and the proposed smoothed ranked based $r_{sa}$ estimator are used for estimating the MSE values. The Figure 2 shows the result of the bivariate normal distribution which is defined below.

$$f_{XY}(x,y) = \frac{1}{2\pi\sqrt{1-\rho^2}} \exp\left\{-\frac{1}{2(1-\rho^2)}[x^2 - 2\rho xy + y^2]\right\}$$

where $-1 \leq \rho \leq 1$.

The Figure 3 is generated from the bivariate exponential distribution, also known as Farlie-Gumbel-Morgenstern distribution, is given by

$$F(x,y) = F_1(x)F_2(y)[1 + \rho(1 - F_1(x))(1 - F_2(y))] \#(39)$$

for $x \geq 0$ and $y \geq 0$ and the marginal distribution functions $F_1$ and $F_2$ are exponential with scale parameters $\theta_1$ and $\theta_2$ and correlation parameter $\rho$ where $-1 \leq \rho \leq 1$. It is important to point out that the marginal distributions of X and Y are exponential with parameters $\alpha$ and $\beta$.

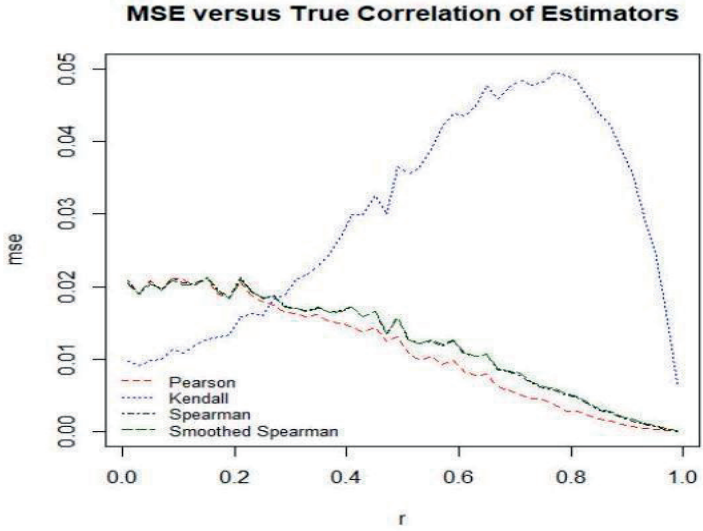

**Figure 2:** Estimation of MSEs of $\rho$ using Bivariate Normal Distribution

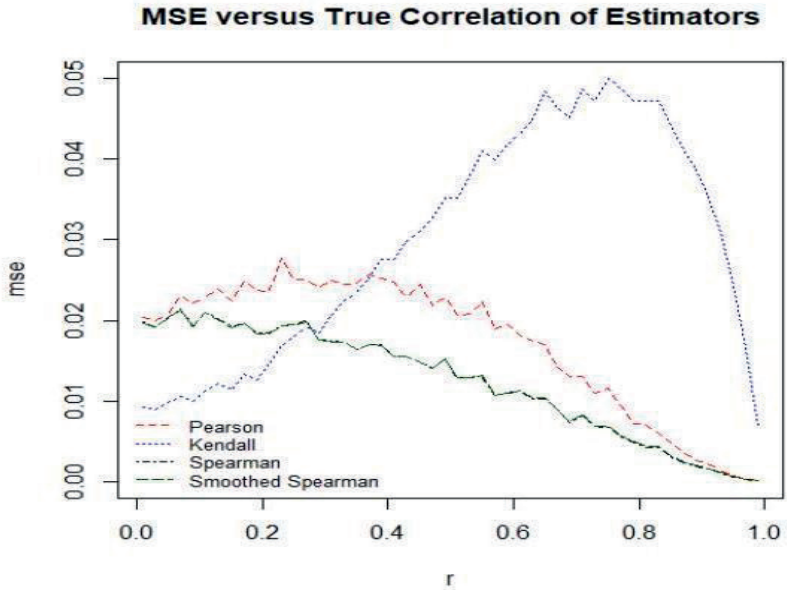

**Figure 3:** Estimation of MSEs of $\rho$ using Bivariate Exponential Distribution

As it can be seen from the Figure 2 under the assumption that X and Y random samples are coming from a bivariate normal distribution, Kendall's $r_k$ gives lower MSEs when about $\rho < 0.30$ but it worsens when $\rho$ approaches 1. On the other hand, Pearson's $r_p$ performs the best overall, gives minimun MSEs when $\rho$ approaches 1. Moreover, the proposed smoothed rank and Spearman estimates MSE estimates exactly overlaps. Therefore, if we compare the all estimators, Pearson's $r_p$ performs the best under the bivariate normality whereas Spearman and the smoothed ranked estimates perform similarly and looks very stable. On the other hand, Kendall's $r_k$ is the worst performer as shown by the Figure 2. As the Figure 3 shows that if we assume X and Y are coming from a bivariate exponential distribution, then proposed smoothed rank and Spearman's correlation estimates performs the best overall. Pearson's MSE values are higher than Spearman and the smoothed rank based estimates under bivariate exponential case. This is expected since Pearson known to be a good estimator under normality or linear relationships. Kendall's $r_k$ shows a fluctuating MSEs as shown previously under the bivariate normal case.

In the second part of the simulation, the relative efficiencies of the smoothed ranked $r_{sa}$ with respected to the other estimators are compared based on the ratios of the estimated MSEs. The results are presented in the Table 1. The most obvious result is that the proposed smoothed rank's $r_{sa}$ estimator performs very similar to the Spearmans's $r_s$ method since the relative efficiencies are very close to 1 which is expected since we saw that similarity on the Figure 2. Pearson shows superiority over the other estimators if the underlying distribution is bivariate normal since the relative efficiency results as $\rho$ goes to 1. Between smoothed and Kendall's estimator, Smoothed rank based estimator has increasing relative efficiencies as $\rho$ increases.

**Table 1:** Relative efficiency rates of the $\rho$ estimators basen on MSEs under bivariate normal

| Correlation($\rho$) | Pears-Smoot | Kendal-Smoot | Spear-Smoot | Pears-Spear |
|---|---|---|---|---|
| 0 | 0.9931 | 0.4690 | 1.0044 | 0.9888 |
| 0.25 | 0.9758 | 0.9021 | 1.0053 | 0.9707 |
| 0.50 | 0.8644 | 2.4973 | 1.0074 | 0.8581 |
| 0.75 | 0.6583 | 7.6484 | 1.0092 | 0.6523 |
| 0.95 | 0.3530 | 38.7826 | 0.9981 | 0.3537 |

Above simulation is repeated under the bivariate exponential distribution condition and results are presented in Table 2.

**Table 2:** Relative efficiency rates of the $\rho$ estimators basen on MSEs under bivariate exponential

| Correlation($\rho$) | Pears-Smoot | Kendal-Smoot | Spear-Smoot | Pears-Spear |
|---|---|---|---|---|
| 0 | 1.0064 | 0.4669 | 1.0024 | 1.0040 |
| 0.25 | 1.3173 | 0.9166 | 1.0057 | 1.3098 |
| 0.50 | 1.6154 | 2.5556 | 1.0065 | 1.6050 |
| 0.75 | 1.7028 | 7.8280 | 1.0131 | 1.6807 |
| 0.95 | 1.1578 | 40.9222 | 1.0479 | 1.1049 |

As we see Table 2, Smoothed rank's $r_{sa}$ and Spearman's $r_s$ performs equally as $\rho$ approaches 1. Both are better than the Pearson's estimator especially when $\rho$ is around $0.50 - 0.75$. The proposed smoothed rank based estimator performs better than the Kendall since relative efficiencies goes as high as 40 as $\rho$ approaches 1.

## 6 CONCLUSIONS

The Spearman's rank correlation coefficient ($\rho$) is a nonparametric measure that assesses the strength and direction of association between two ranked variables. Traditionally, $\rho$ is computed as

$$\rho = 1 - \frac{6 \sum D_i^2}{n(n^2 - 1)} \#(40)$$

where $D_i$ is the difference between the ranks of corresponding variables $x$ and $y$, and $n$ is the number of observations. However, this method can be sensitive to the exact rank assignments, particularly when there are ties or the noise is present in the data. To address these issues, smoothed of ranks can be employed in the Wilcoxon's linear score function which would enhance the robustness and sensitivity of Spearman's correlation under several conditions such as existence of ties or extreme monotonic associations.

The advantages of using smoothed ranks in Spearman's correlation are manifold. First, they provide a more nuanced understanding of the relationship between variables, particularly in the presence of noisy or tied data. Second, the approach is more robust to outliers and non-linear relationships. Third, it allows for a more flexible interpretation of ranks, accommodating the specific characteristics and distributions of the data. The proposed method can yield more accurate and meaningful insights into the underlying monotonic relationships, especially in complex data sets where traditional methods might not work the best. Finally, one of the future research objectives would be implementation of adaptive score functions that would address non-monotone associations when Spearman's correlation might not be an ideal method.

VI. International Applied Statistics Congress (UYIK-2025)
Ankara / Türkiye, May 14-16, 2025

**Acknowledgement**

We would like to send our special thanks to the UYIK-2025 Scientific Committee for their valuable feedback to improve this article.

**Conflict of Interest**

Authors declared that there is no conflict of interest in preparation of this article with used of any data sets.